\documentclass[8.5pt,twoside,twocolumn]{article}
\usepackage[super,sort&compress,comma]{natbib}
\usepackage[french, ngerman,english]{babel}

\usepackage{times,mathptm}
\usepackage{sectsty}
\usepackage{balance}
\usepackage{bm}

\usepackage{textcomp}

\usepackage{amsmath}
\usepackage{graphicx} 
\usepackage[format=plain,justification=raggedright,singlelinecheck=false,font=small,labelfont=bf,labelsep=space]{caption}
\usepackage{fancyhdr}
 \usepackage[version=3]{mhchem}
 \usepackage[table]{xcolor}
 \usepackage{units}

\usepackage{ulem}
\usepackage{amssymb,amsmath}

\begin{document}

\thispagestyle{plain}
\fancypagestyle{plain}{
\renewcommand{\headrulewidth}{1pt}}
\renewcommand{\thefootnote}{\fnsymbol{footnote}}
\renewcommand\footnoterule{\vspace*{1pt}%
\hrule width 3.4in height 0.4pt \vspace*{5pt}}
\setcounter{secnumdepth}{5}

\newcommand\Rey{\mbox{\textit{Re}}}  
\newcommand\Wi{\mbox{\textit{Wi}}} 
\newcommand\El{\mbox{\textit{El}}}  
\newcommand\De{\mbox{\textit{De}}}  

\newcommand\s{\dot{\gamma}}

\newcommand\al{\textit{et~al.}\ }

\def\mum{\nobreak\mbox{$\;$\textnormal{\textmu m}}}
\def\mlmin{\nobreak\mbox{$\;$\textnormal{\textmu l/min}}}
\def\m{\nobreak\mbox{$\;$\textnormal{\textmu}}}

\makeatletter
\def\subsubsection{\@startsection{subsubsection}{3}{10pt}{-1.25ex plus -1ex minus -.1ex}{0ex plus 0ex}{\normalsize\bf}}
\def\paragraph{\@startsection{paragraph}{4}{10pt}{-1.25ex plus -1ex minus -.1ex}{0ex plus 0ex}{\normalsize\textit}}
\renewcommand\@biblabel[1]{#1}
\renewcommand\@makefntext[1]%
{\noindent\makebox[0pt][r]{\@thefnmark\,}#1}
\makeatother
\renewcommand{\figurename}{\small{Fig.}~}
\sectionfont{\large}
\subsectionfont{\normalsize}

\newcommand{\comA}[1]{\textcolor{blue}{\it{#1}}}
\newcommand{\comC}[1]{{\color{red}{\it{#1}}}}

\fancyfoot{}
\fancyfoot[RO]{\footnotesize{\sffamily{1--\pageref{LastPage} ~\textbar  \hspace{2pt}\thepage}}}
\fancyfoot[LE]{\footnotesize{\sffamily{\thepage~\textbar\hspace{3.45cm} 1--\pageref{LastPage}}}}
\fancyhead{}
\renewcommand{\headrulewidth}{1pt}
\renewcommand{\footrulewidth}{1pt}
\setlength{\arrayrulewidth}{1pt}
\setlength{\columnsep}{6.5mm}
\setlength\bibsep{1pt}

\twocolumn[
  \begin{@twocolumnfalse}
\noindent\LARGE{\textbf{Serpentine channels: micro -- rheometers for fluid relaxation times$^\dag$}}
\vspace{0.6cm}

\noindent\large{\textbf{Josephine Zilz\textit{$^{a}$}, Christof Sch{\"a}fer\textit{$^{b}$}, Christian Wagner\textit{$^{b}$}, Robert J. Poole\textit{$^{c}$}, Manuel A. Alves\textit{$^{d}$}and
Anke Lindner\textit{$^{a}$}}}\vspace{0.5cm}

\noindent\textit{\small{\textbf{Received Xth XXXXXXXXXX 20XX, Accepted Xth XXXXXXXXX 20XX\newline
First published on the web Xth XXXXXXXXXX 200X}}}

\noindent \textbf{\small{DOI: 10.1039/b000000x}}
\vspace{0.6cm}

\noindent \normalsize{We propose a novel device capable of measuring the relaxation time of viscoelastic fluids as small as 1\,ms.  In contrast to most rheometers, which by their very nature are concerned with producing viscometric or nearly-viscometric flows, here we make use of an elastic instability which occurs in the flow of viscoelastic fluids with curved streamlines.  To calibrate the rheometer we combine simple scaling arguments with relaxation times obtained from first normal-stress difference data measured in a classical shear rheometer.  As an additional check we also compare these relaxation times to those obtained from Zimm theory and good agreement is observed. Once calibrated, we show how the serpentine rheometer can be used to access smaller polymer concentrations and lower solvent viscosities where classical measurements become difficult or impossible to use due to inertial and/or resolution limitations. In the absence of calibration the serpentine channel can still be a very useful comparative or index device.}
\vspace{0.5cm}
 \end{@twocolumnfalse}
  ]

\footnotetext{\dag~Electronic Supplementary Information (ESI) available: [details of any supplementary information available should be included here]. See DOI: 10.1039/b000000x/}
\footnotetext{\textit{$^{a}$~PMMH, ESPCI, UPMC, Univ. Paris-Diderot, CNRS UMR 7636 10, rue Vauquelin F-75231 Paris Cedex 05, France}}
\footnotetext{\textit{$^{b}$~Experimentalphysik, Saarland University - D-66123, Saarbr{\"u}cken,  Germany.}}
\footnotetext{\textit{$^{c}$~School of Engineering, University of Liverpool, Brownlow Hill, Liverpool, L69 3GH, United Kingdom.}}
\footnotetext{\textit{$^{d}$~Departmento de Engenharia Quimica, Faculdade de Engenharia da Universidade do Porto, Rua Dr. Roberto Frias, 4200-465 Porto, Portugal. }}

\section{Introduction}

\begin{figure}[h!]
	\centering
		\includegraphics[width=0.48\textwidth]{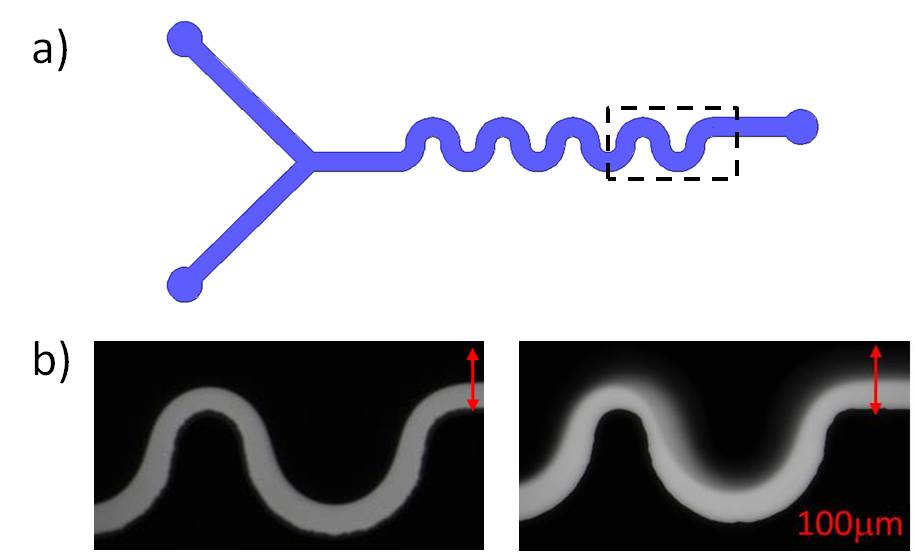}
	\caption{a) Schematic of the microfluidic serpentine channel b) Snapshots from the experiments showing the instability onset. Solutions of PEO (4Mio-1 $c$ = 125\,ppm; $\eta_s$ = 4.9\,mPa\,s) are injected into a microchannel ($W$ = $H$ \(\sim\) 100$\mum$; $R$ = 200$\mum$) via two inlets; only one stream contains fluorescent dye and is visible on the snapshot. Left hand side: flow rate $Q=20\mlmin$ (below instability onset), right hand side: $Q=40\mlmin$ (above instability onset).}
	\label{fig:serpentine_channel}
\end{figure}

Polymer solutions of long and flexible polymers are known to exhibit striking non-Newtonian properties even at very small concentrations \cite{Bird1987}.  For example, in turbulent pipe or channel flow the friction factor (or drag) may be significantly reduced by adding a polymer at concentrations as low as a few ppm \cite{Virk1975} (parts per million in weight); such fluids are also used in enhanced oil recovery applications \cite{Jones1989}.  Measuring their rheological features is a challenging task and classical rheometry is often at its limits when determining for example relaxation times of such dilute polymer solutions\cite{Lindner2003}.

Here we develop a microfluidic rheometer with a complex flow geometry to overcome these difficulties.  Although a number of microfluidic rheometers have been proposed, most of these devices are restricted to measurements of shear viscosity \cite{Guillot2006,Guillot2008, Pipe2008,Chev2008,Nghe2010,Livak2013, Berthet2011} although devices which attempt to estimate extensional viscosity \cite{Pipe2009,Oliveira2008,Nelson2011,Haward2012,GR2013} and dynamic properties \cite{Chris2010} have also been proposed. In contrast to these previous microfluidic devices, in the current study we make use of an elastic instability\cite{Larson1990,Groisman2004a,Shaqfeh1996,McKinley1996,Khom2000,Grois2004,Poole2007,Morozov2007,Bon2011}, that occurs in flows of viscoelastic fluids with curved streamlines even in the absence of inertia\cite{Shaqfeh1996,McKinley1996}. The threshold of instability depends on the curvature of the flow and the fluid elasticity \cite{Pakdel1996}, described by the Weissenberg number. Typically, viscoelastic effects will be observed when the product of a fluid relaxation time $\lambda$ and a characteristic shear rate reaches order one: thus for fluids with $\lambda$ on the order of milliseconds, shear rates on the order of $10^3$\,s$^{-1}$ are required to access such viscoelastic effects. The use of a microfluidic device enables high shear rates to be obtained and thus strong viscoelastic effects (corresponding to large Weissenberg numbers) to be observed while keeping inertial effects, and hence the Reynolds number, small.

 We have recently investigated the flow in a serpentine micro-channel to elucidate the scaling of the instability threshold with the flow curvature using a dilute polymer solution \cite{Zilz2012}. We have shown that the instability is very sensitive to even small normal-stress differences and can thus be used to detect their occurrence. We can now combine our precise knowledge regarding the dependence of the instability onset on the flow curvature with a precise knowledge of the rheological properties of a calibrating fluid to quantitatively measure relaxation times. To do so, we initially calibrate the serpentine rheometer using classical shear rheometry in the range of parameters accessible by this technology. The serpentine rheometer can then be used with fluids of smaller concentrations or lower solvent viscosities where classical rheometry techniques become difficult either due to inertial instabilities or instrument resolution issues. Even when a precise calibration is not possible, the serpentine channel can be used as a comparative rheometer, to compare the rheological properties of two given fluids. Finally we propose methods to fully integrate the serpentine channel into a microfluidic lap-on-a-chip device capable of measuring both shear viscosity and fluid relaxation time.

\section{Scaling of the onset of elastic instability in a serpentine channel}

Pakdel and McKinley \citep{Pakdel1996, McKinley1996} proposed a simple dimensionless criterion that must be exceeded for the onset of purely-elastic instability, combining the curvature of the flow and the tensile stress $\tau_{11}$ acting along the streamlines in the following form:

\begin{equation}
\left[\frac{\tau_{11}}{\eta \s}\frac{\lambda u}{{\cal
R}}\right]\ge M^2_{crit}
\label{equ:PakMc}
\end{equation}\

with ${\cal R}$,  $u$ and $\s$
representing the local streamline radius of curvature, velocity magnitude and shear rate, respectively. $\tau_{11}$ represents the local streamwise
normal-stress and $\eta\s$ the local shear stress, with $\eta$ being the shear viscosity. The ratio $\tau_{11}/\eta \s$ thus represents a local Weissenberg number $\Wi$, comparing normal stresses to shear stresses and $\lambda u/\cal R$
 compares a typical distance over which a polymer relaxes to the radius of curvature (or can be viewed as a local Deborah number).

\begin{figure}[h]
	\centering
		\includegraphics[height=6.5cm]{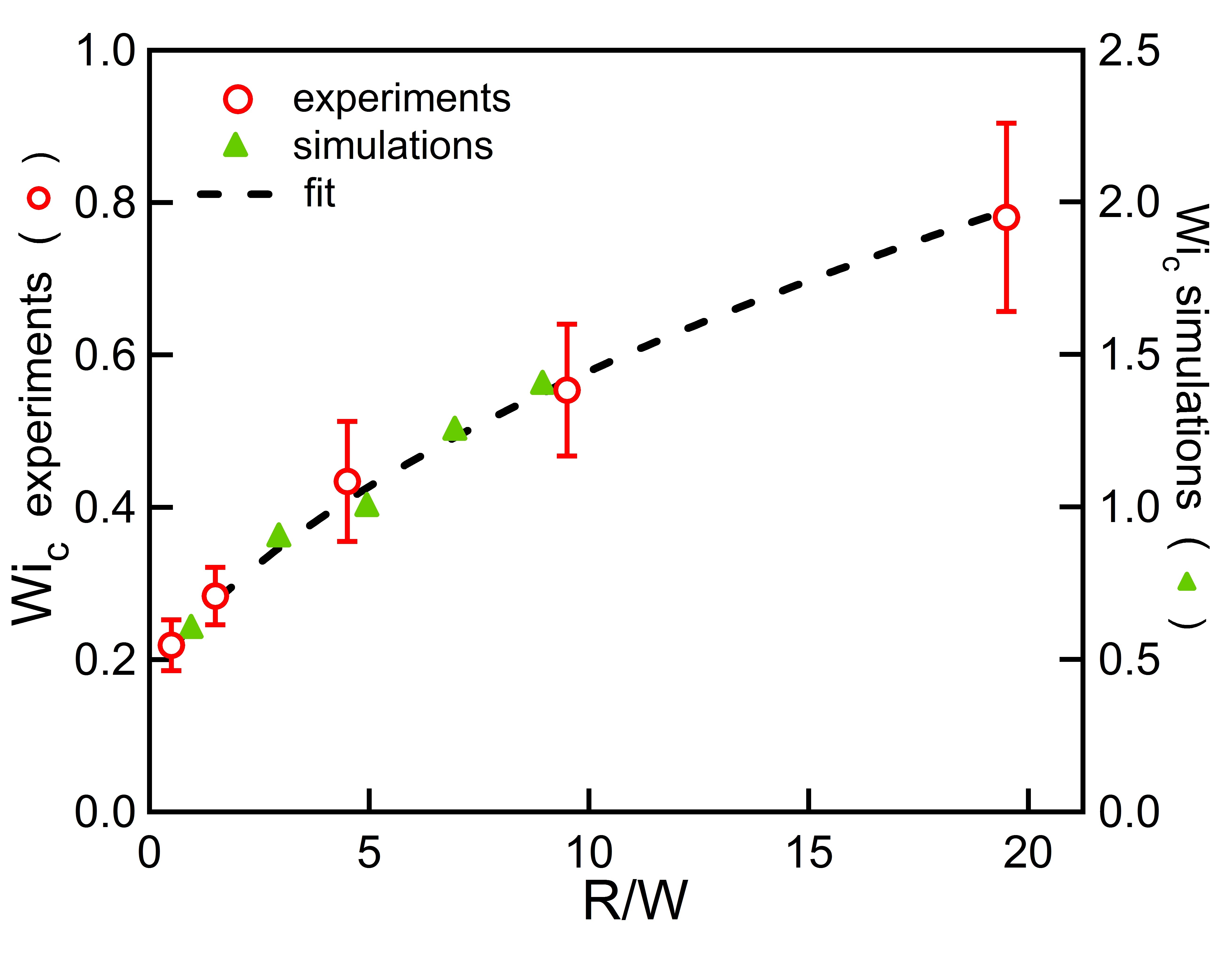}
	\caption{Geometric scaling of the instability onset. The green triangles are numerical results and the red circles are results from experiments. The dotted line is a fit to equation \eqref{equ:geometric_scaling}. Data from Zilz \al\cite{Zilz2012}.}
	\label{fig:geometric_scaling}
\end{figure}

We have recently elucidated the geometrical scaling for the onset of elastic-flow instability in a serpentine channel  by adapting the Pakdel-McKinley criterium to the specific flow geometry \cite{Zilz2012}. The serpentine channel is composed of a series of circular half-loops of alternating curvature of constant width $W$, height $H$ and inner radius $R$, as shown schematically in Fig.\,\ref{fig:serpentine_channel}a.  For reasons of simplicity in our analysis for the shear-dominated flow in the serpentine channel we used the upper-convected Maxwell (UCM) model, neglecting the solvent viscosity $\eta_s$ contribution. The total viscosity $\eta$ is thus simply equal to the polymer viscosity $\eta_p$ (i.e. $\eta=\eta_p$) and the normal-stress is approximated as $\tau_{11}=2\eta_p\lambda\s^2$. In this case the ratio $\tau_{11}/\eta \s$ becomes equal to $2\lambda\s$ corresponding to twice the Weissenberg number.
A simple analysis based on the Pakdel-McKinley criterion (eqn.\ \eqref{equ:PakMc}) showed that the critical Weissenberg number at instability onset $\Wi_c$ can be written as a square root dependence on the normalized radius $R/W$ with a small off-set at small radii. For a channel with square cross section,  our numerical results\cite{Zilz2012} (see Fig.\,\ref{fig:geometric_scaling}) are best described in the following form:

\begin{equation}
\Wi_c=C\sqrt{1+\frac{R}{W}} \;.
\label{equ:geometric_scaling}
\end{equation}

 Note that the numerical value for the offset at small radii found from the numerical results differs slightly from the theoretical prediction given in \cite{Zilz2012} as the flow asymmetry occurring at strong curvature is not captured by our theoretical model, as has already been pointed out in \cite{Zilz2012}.  The predicted scaling is in excellent agreement with experimental observations, as shown in Fig.\,\ref{fig:geometric_scaling}. Here we want to go further and not only obtain the scaling of the instability onset with the flow geometry, but reach a quantitative prediction of the instability threshold. To do so, one first has to take the solvent viscosity contribution into account, which cannot be neglected for the dilute polymer solutions we use. Using an appropriate model, as for example the Oldroyd-B model\cite{Bird1987} to describe the polymer rheology so that the total viscosity $\eta$ is comprised of both a polymer contribution $\eta_p$ and a solvent contribution $\eta_s$, i.e. $\eta=\eta_p+\eta_s$, the scaling for the instability onset has to be corrected\cite{McKinley1996,Poole2013} using $\tau_{11}/\eta\s=2(\eta_p/\eta)\lambda \s $. Doing so and then rewriting a modified form of equation \eqref{equ:geometric_scaling} to obtain the critical shear rate one obtains:

\begin{equation}
\s_c=\frac{C}{\lambda} \sqrt{\frac{\eta}{\eta_p}} \sqrt{1+\frac{R}{W}}
\label{equ:geometric_scaling_visc}
\end{equation}

where $(C/\lambda) \sqrt{\eta/\eta_p}$ can be identified as the slope from a plot of the critical shear rate $\s_c$ {\it vs} $\sqrt{1+R/W}$. At this juncture it is also useful to define a parameter $a=\lambda/C$. To be able to make a quantitative prediction of the relaxation time $\lambda$ from measurements of the critical shear rate $\s_c$ one thus needs a calibration experiment to determine $C$ and the ratio of the polymer to the solvent viscosity. We note also that the serpentine rheometer strictly only allows for a quantitative measurement of the polymer relaxation time, as long as the rheology of the solution is such that the ratio between the normal stresses and the shear stresses is proportional to $\Wi$.

\begin{figure}[h]
	\centering
		\includegraphics[height=6.5cm]{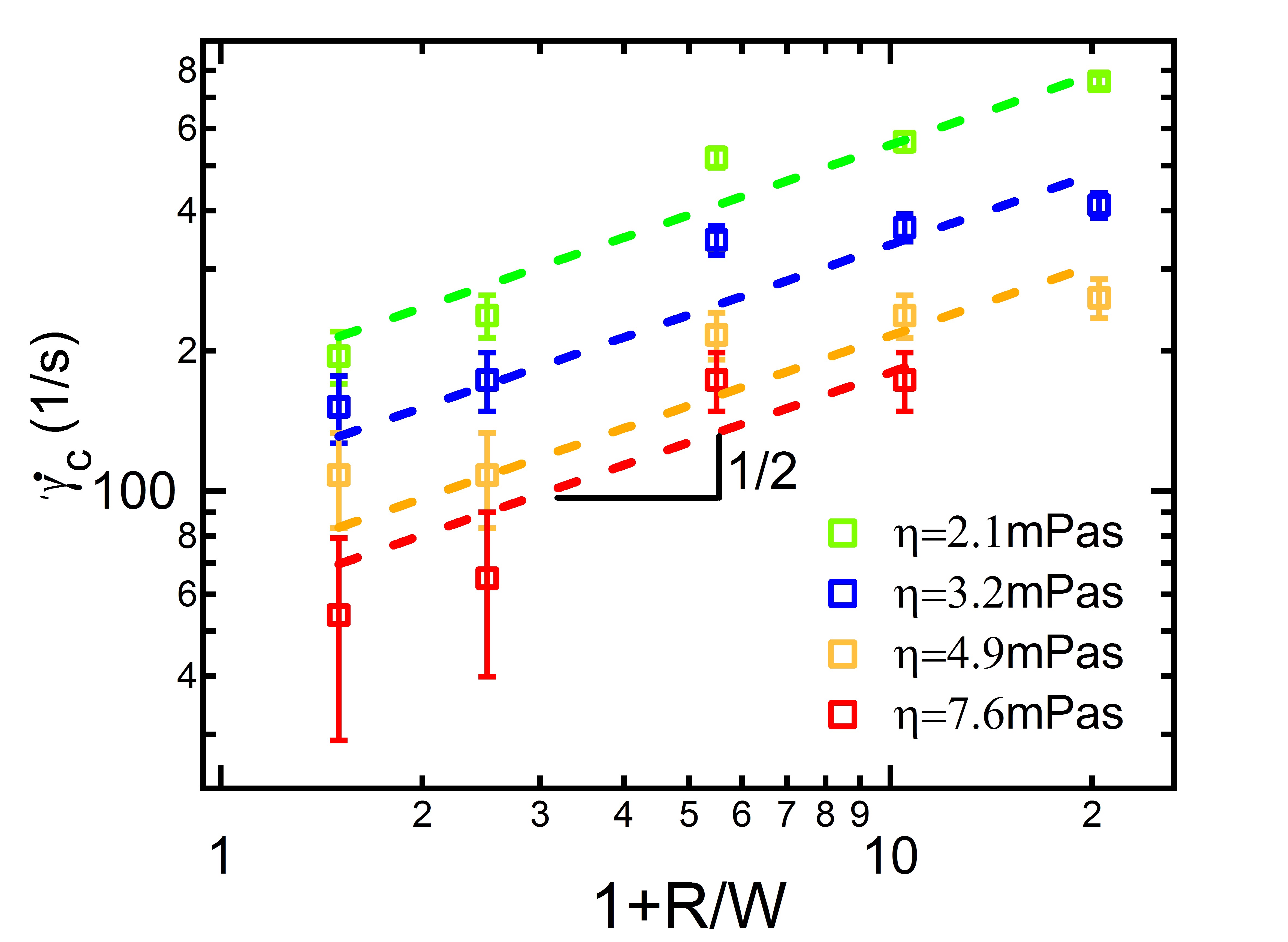}
	\caption{Critical shear rate $\s_c$ as a function of the normalized radius $1+R/W$ for solutions of PEO 2Mio at a concentration of 400\,ppm for different solvent viscosities $\eta_s$. The dotted lines represent fits to the data using equation \eqref{equ:geometric_scaling_visc}. Each experiment was repeated at least two times using fresh polymer solutions and the average value is shown together with the error bars.}
	\label{fig:fits_serpentine channel}
\end{figure}

\section{Experimental}
\subsection{The polymer solutions}

Solutions of the flexible polymer polyethylene oxide (PEO), supplied by Sigma Aldrich, with nominal
molecular weight $M_w = 2\times10^6$\,g/mol and two different batches of PEO with nominal molecular weight $4\times10^6$\,g/mol, at concentrations ranging from 125\,ppm to 500\,ppm (w/w), were used in water/glycerol mixtures. In the following the different polymers will be referred to as 2Mio and 4Mio-1 and 4Mio-2, respectively. The overlap
concentration for 2Mio is $c^* \simeq 860$\,ppm\cite{Rodd2007} and for the 4Mio, estimated using the equations provided in Rodd et al\cite{Rodd2007}, is $c^* \simeq 550$\,ppm and therefore the solutions are dilute in all cases ($c/c^*< 1$). The solvent viscosity $\eta_s$ varied from  $\eta_s=1$\,mPa\,{}s for pure water to 10.7\,mPa\,s, at 20\,$^\circ$C, for varying concentrations of glycerol. All polymer solutions and concentrations used are summarized in Table \ref{tab:table}.

\subsection{The serpentine channel}

The serpentine channels used in this work consist of a series of 8 half-loops of width $W=100\mum$,
height $H=80\mum$ and varying inner radius $R$. The number of loops and the
geometry of the inlets of the serpentine channels used in this study are
represented in Fig.\ \ref{fig:serpentine_channel} and have been described in detail in Zilz \al \cite{Zilz2012}. Experiments performed with varying number of loops ($N$) confirmed that the results presented here are independent of the exact number of loops provided $2 < N < 15$. The channels are made from PDMS, but due to the small viscosity of the polymer solutions used the applied pressure remained sufficiently small to avoid deformation of the channels. The solutions are supplied to the microchannel via two inlets, one stream of which is fluorescently dyed. The light grey area visible in the snapshots of Fig.\,\ref{fig:serpentine_channel}b) shows the location of the fluorescently labeled fluid. Its width variation along the streamwise direction shows the slight asymmetry of the flow field along the flow path due to local flow acceleration in the curved geometry. Note that the channel width is constant over the whole length of the channel. Fig.\,\ref{fig:serpentine_channel}b) shows a stable flow situation on the left panel and an unstable flow situation on the right panel. In this way the stability of the flow can easily be visually assessed and will always be monitored at the last loop. The time-dependent flow is easily identifiable in the real-time flow visualization. The flow rate $Q$ was varied from 1 to 50\,$\mlmin$, and was imposed via a syringe pump (PHD 2000, Harvard apparatus). The Reynolds number $\Rey$ is defined as $\Rey=\rho U W/\eta$, with $\rho$ representing the density of the fluid and $U=Q/WH$ the average velocity in the channel. The maximum $\Rey$, corresponding to the highest flowrate and lowest viscosity solution, never exceeded 5. The flow is visualized using an inverted microscope (Axio Observer,
Zeiss) coupled to a CCD camera (PixeLink). Starting with the lowest flow rate,
$Q$ is then gradually increased. After each change in $Q$ a
sufficiently long time is allowed to achieve steady-state flow conditions (on
the order of 10 minutes per step). The onset of fluctuations in the flow defines the onset of the time-dependent elastic
instability, and the critical flow rate $Q_c$ is determined. From the critical flow rate we obtain the critical average shear rate $\s_c$, defined as $\s_c=Q_c/(W^2H)$.

As an example Fig.\,\ref{fig:fits_serpentine channel} illustrates the results obtained for PEO 2Mio at a concentration of 400\,ppm. Similar experiments (not shown) have been performed at different concentrations and for different molecular weights as given in Table \ref{tab:table}. In parallel to each experiment the solvent viscosity and the ratio between polymer and solvent viscosity $\eta_p/\eta_s$ was determined using a Ubbelohde capillary viscometer. In this way it was possible to correct for small changes in temperature that occurred in the laboratory (typically between 20\,$^\circ$C and 23\,$^\circ$C).

\begin{table}
\begin{center}
  \caption{Polymer solutions used in the serpentine channel and fit parameters of $\lambda=A\eta_s^{0.9}$.
   \label{tab:table}}
  \begin{tabular}{|c|c|c|}\hline
  \multicolumn{3}{|c|}{Polymer solutions used in the serpentine channel}\\
  \hline
    $M_w$ &concentration & $\eta_p/\eta_s$  \\ \hline
   2Mio & 125-500\,ppm & $7\%-39\%$  \\
   4Mio-1 & 400\,ppm &  $34\%$ \\
  4Mio-2 & 400\,ppm & $34\%$  \\
  \hline
    \multicolumn{3}{|c|}{Fit parameters of $\lambda=A\eta_s^{0.9}$ from classical rheometry}\\ \hline
      $M_w$ &concentration & $A$ ms/(mPa\,s)$^{0.9}$  \\ \hline
   2Mio & 400\,ppm & $0.25\pm 0.02$  \\
   4Mio-1 & 400\,ppm &  $0.59\pm0.02$ \\
  4Mio-2 & 400\,ppm & $0.95\pm0.04$  \\
  \hline
  \end{tabular}
  \end{center}
\end{table}

\subsection{The classical rotational rheometer}

\begin{figure}[h]
	\centering
\includegraphics[width=8.5cm]{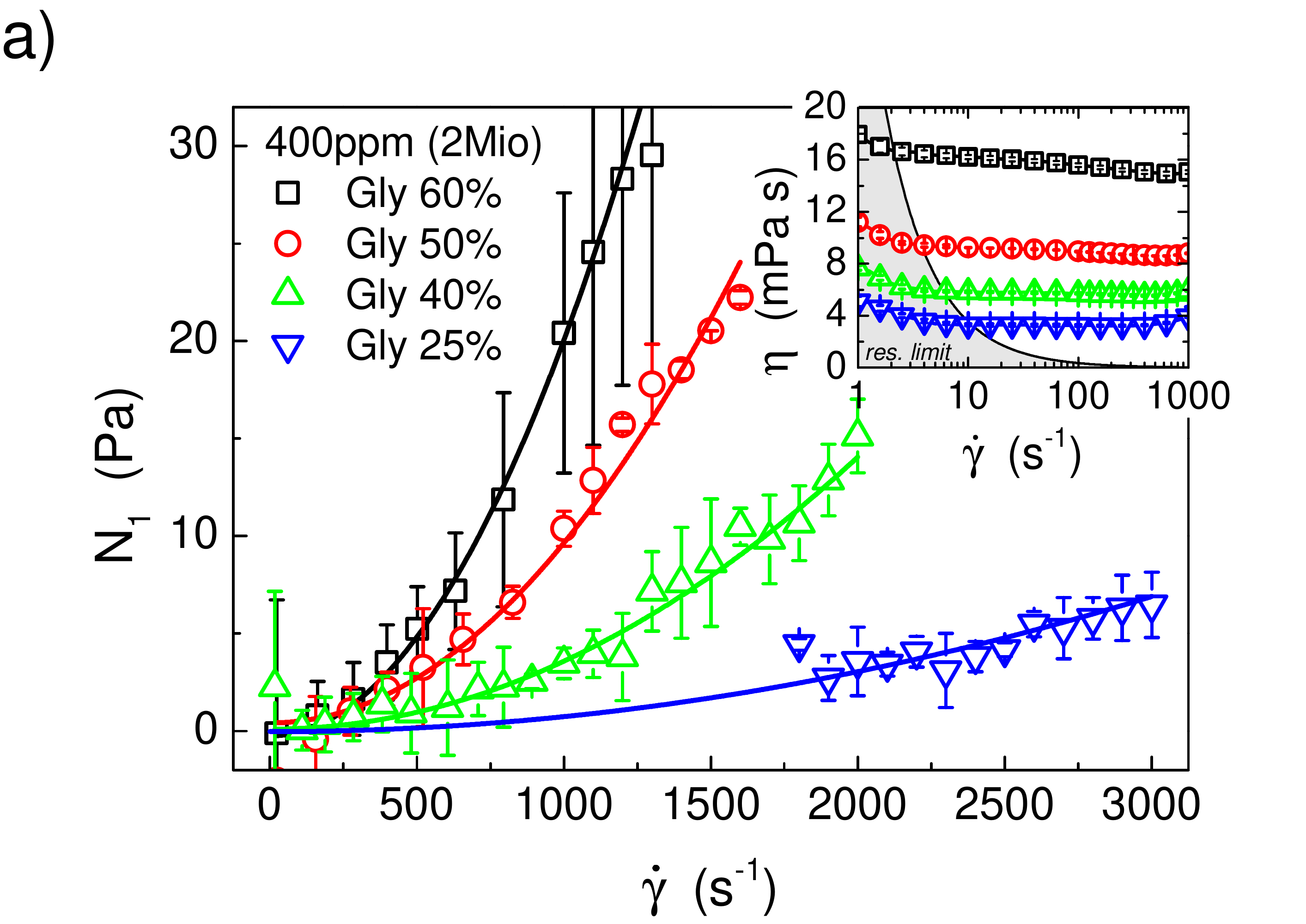}\\
		\vspace{-0.25cm}
		\includegraphics[width=8cm]{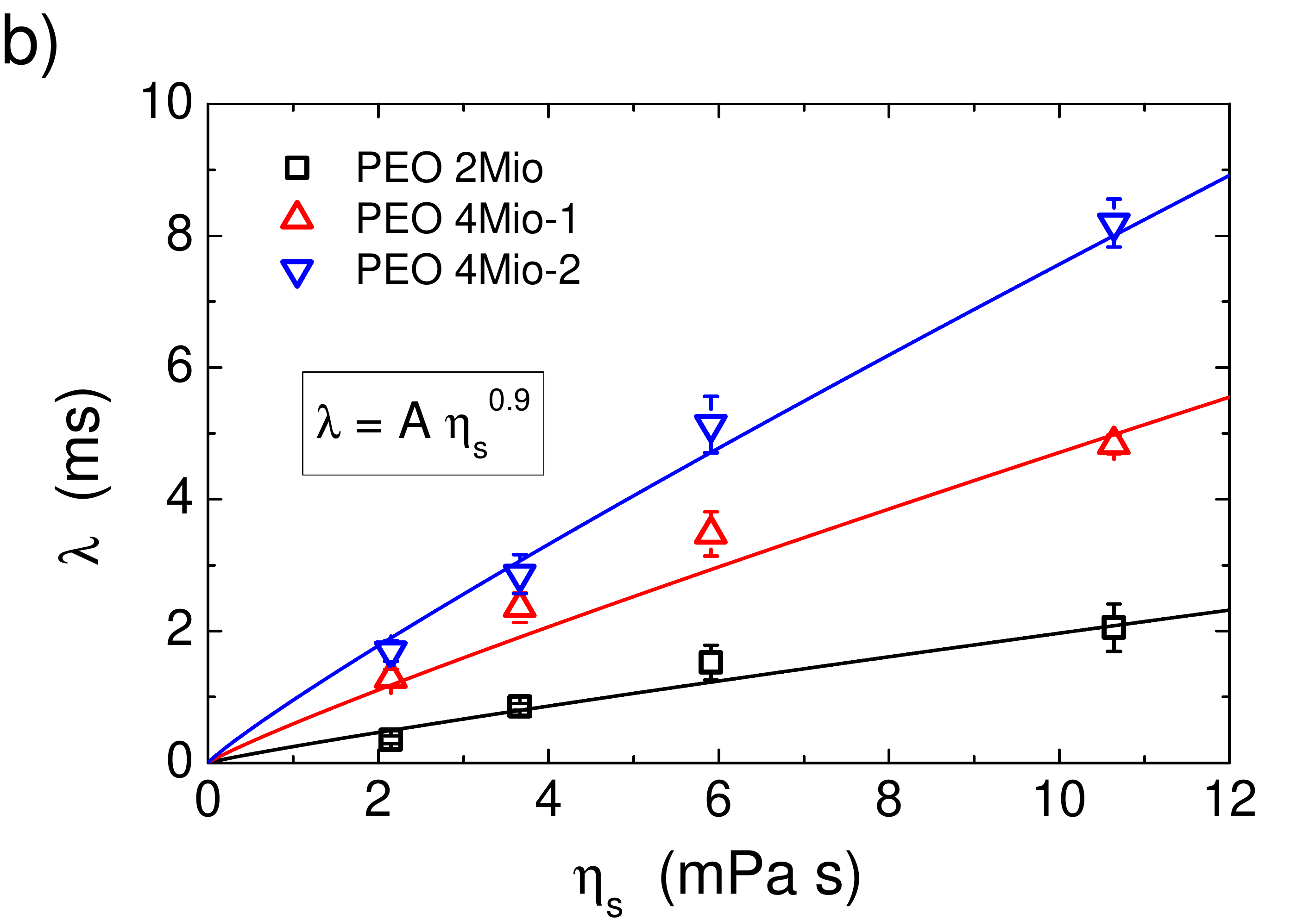}
	\caption{a) Normal-stress $N_1$ as a function of the shear rate from classical rheometry for PEO 2Mio at 400\,ppm. A quadratic fit to the normal-stress data is also indicated. Inset: shear viscosity $\eta$ as a function of shear rate. Note that for the representation of the data an average over several runs (at least 3) has been plotted. b) $\lambda$ {\it vs} $\eta_s$ from classical shear rheometry at a polymer concentration of 400\,ppm. The fits correspond to $\lambda=A\eta_s^{0.9}$.}
	\label{fig:rheo_SB}
\end{figure}

A commercial rotational shear rheometer (MARS II, Thermo Scientific) in combination with a cone-and-plate geometry (diameter $D=60$\,mm, cone angles $\alpha=2\,^\circ$ and $1\,^\circ$) in shear rate controlled mode was used to measure the viscosity $\eta$ and first normal-stress difference $N_1 = \tau_{11} - \tau_{22}$ simultaneously following the methodology laid out in Zell \al \cite{Zell2010}. In models of dilute polymer solutions $\tau_{22}$ is negligible\cite{Larson1999} in the steady simple shear flow experiments and thus $N_1$ is identical to the streamwise normal-stress $\tau_{11}$. The shear rate was increased in a step-wise protocol from $1$\,s$^{-1}$ up to a maximum of $3000$\,s$^{-1}$, with $15$\,s of equilibration time at each shear rate. The temperature was kept constant at $T=20\pm 0.5\,^\circ$C by using a Haake Phoenix II refrigerated circulator. The normal-stress data of the polymer solutions was corrected by taking into account inertial contributions that can be easily obtained from $N_1$ measurements of the Newtonian solvents. The constant solvent viscosities $\eta_s(\dot\gamma) \equiv \eta_s$ were also measured. The $\Psi_1$ data of the polymer solutions was obtained from quadratic fits to the corrected data for $N_1$ according to $N_1(\s)=\Psi_1\s^2$ within adequate ranges of shear rate. An example of the normal-stress data $N_1$ together with the results for the shear viscosity $\eta$ is shown in Fig.\,\ref{fig:rheo_SB}a) for PEO 2Mio at a concentration of 400\,ppm. A quadratic fit to the normal-stress data is also indicated. Note that for the representation of the data an average over several runs (at least 3) has been plotted. The relaxation time $\lambda$ of the polymer was determined by taking

\begin{equation}
\lambda = \frac{\Psi_1}{2\eta_p(\dot\gamma)} = \frac{\Psi_1}{2[\eta(\dot\gamma)-\eta_s(\dot\gamma)]}\;.
\label{equ:lambda_rheo}
\end{equation}

When calculating the relaxation time we refer to the polymer viscosity at a fixed shear rate of $\dot\gamma = 100$\,s$^{-1}$ and neglect the slight shear-thinning behavior of the polymer solutions. The associated uncertainties $\delta\eta$, $\delta\eta_s$ and $\delta\Psi_1$ (from the statistics of multiple independent measurements) can be interpreted in terms of an estimate of the uncertainty of $\lambda$, i.e.\ $\delta \lambda^2 \lesssim \left(\delta\Psi_1/2\bar\eta_p\right)^2 + \left(\bar\Psi_1/2\bar\eta_p^2\right)^2\left(\delta\eta^2 + \delta\eta_s^2\right)$ with mean values denoted by an overbar. The variations of $\lambda = \bar{\lambda} \pm \delta\lambda$ with solvent viscosity are shown in Fig.\,\ref{fig:rheo_SB}b) and can be described as $\lambda=A\eta_s^{0.9}$ as has been found as the best fit for all three curves presented in Fig.\,\ref{fig:rheo_SB}b). The fit parameters $A$ are shown in Table \ref{tab:table}.

\begin{figure*}[th!]
	\centering
\includegraphics[height=6.5cm]{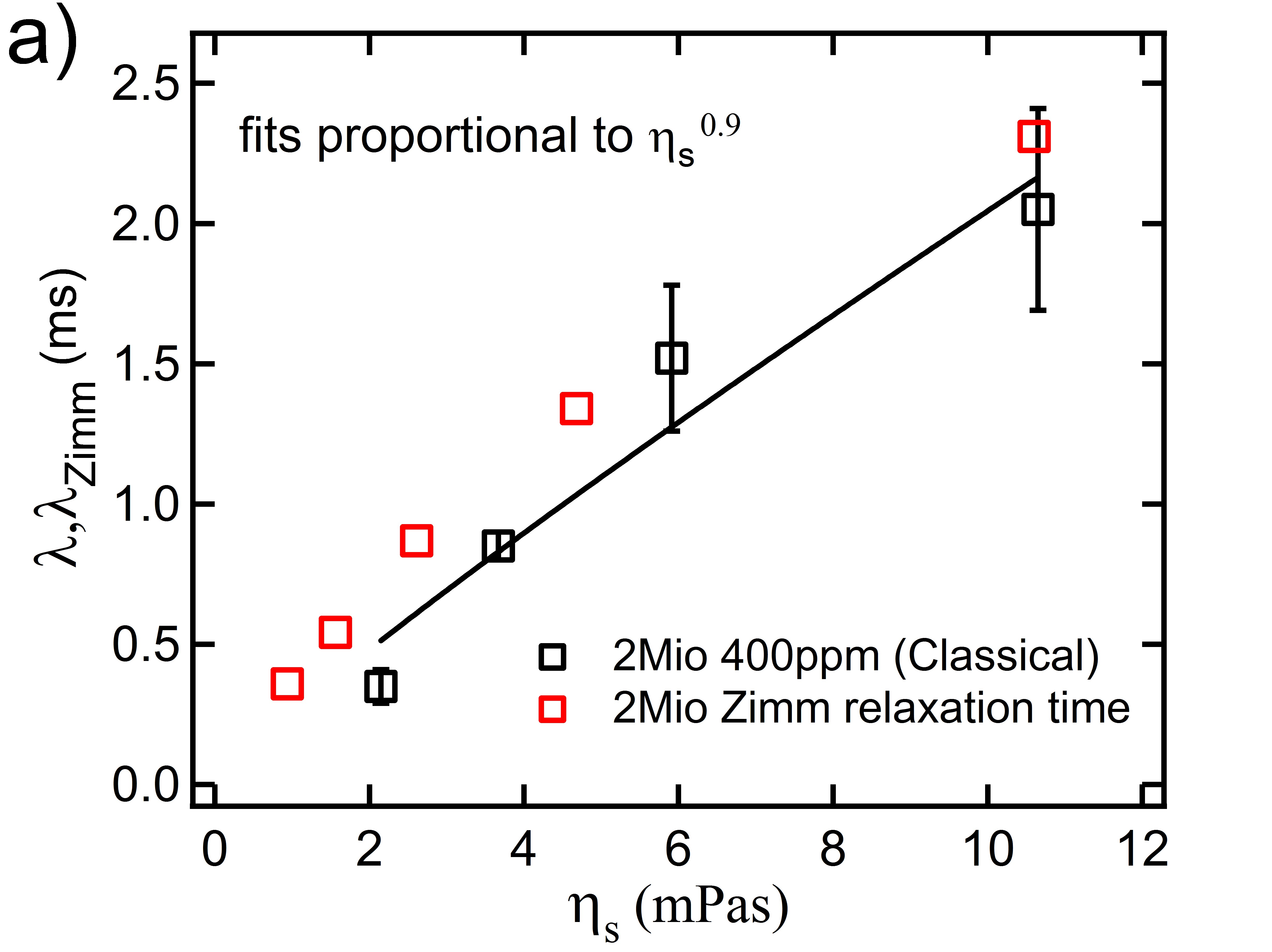}
       \includegraphics[height=6.5cm]{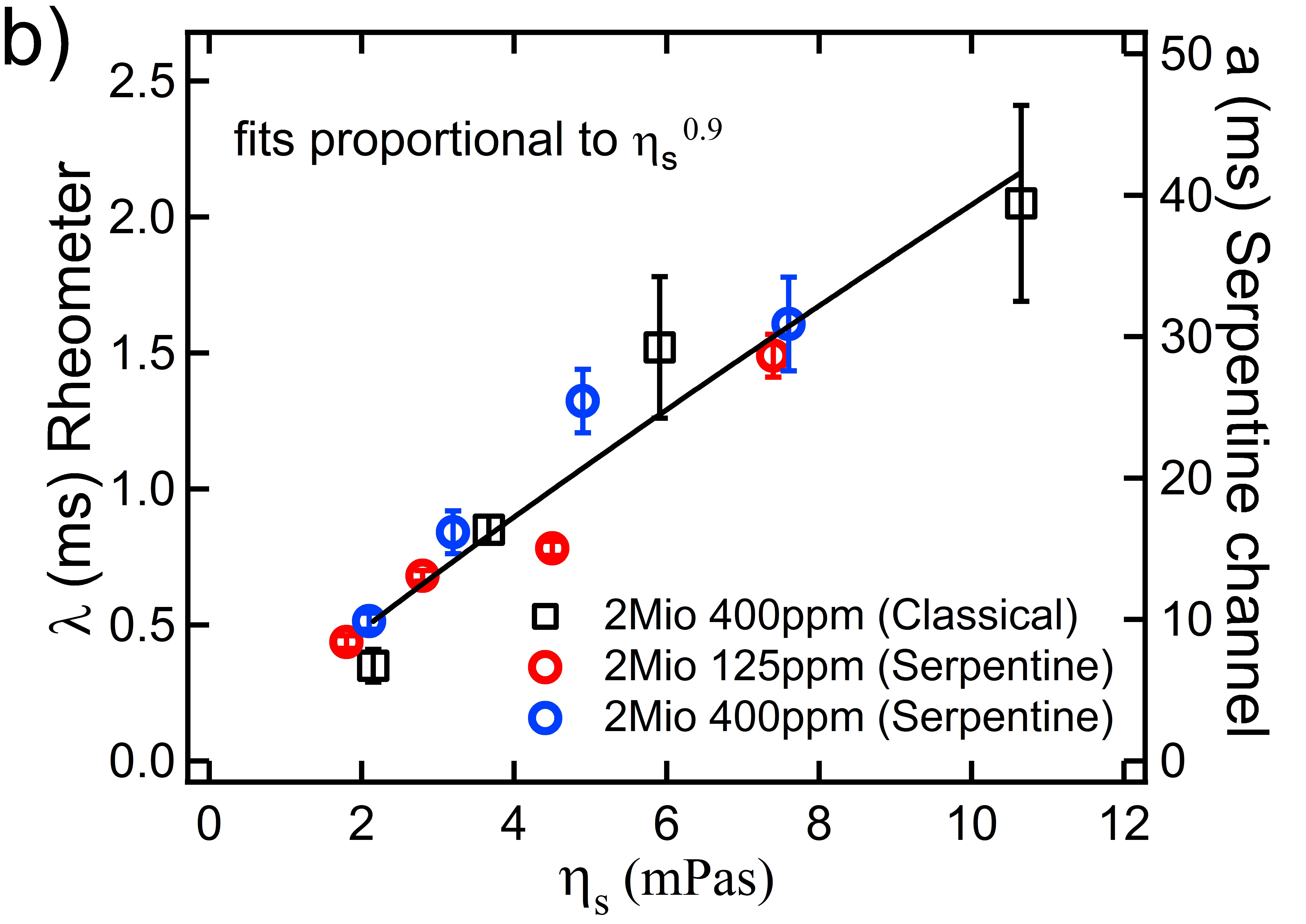}
	\caption{a) Relaxation time from classical shear rheometry (eqn.\ \eqref{equ:lambda_rheo}) compared to Zimm relaxation time\cite{Rodd2005,Rodd2007} (eqn.\ \eqref{equ:Zimm}) b) Relaxation time from classical shear rheometry (eqn.\ \eqref{equ:lambda_rheo}) versus solvent viscosity $\eta_s$ for PEO 2Mio and a concentration of 400ppm (left axis). $a=\lambda/C$ (eqn.\ \eqref{equ:geometric_scaling_visc}) from the serpentine rheometer versus solvent viscosity $\eta_s$ for PEO 2Mio and concentrations of 125\,ppm and 400\,ppm (right axis). The error for the data from the serpentine rheometer corresponds to the error of the fit to $\s_c$ (eqn.\ \eqref{equ:geometric_scaling_visc}), and for the data from the classical rheometer the errors are estimated by an error propagation according to the uncertainties of $\Psi_1$ and $\eta_p$.}
	\label{fig:calibration}
\end{figure*}

For a polymer chain in a good solvent, according to Zimm theory\cite{Larson1999}, the longest relaxation time of a dilute solution can be estimated using
\begin{equation}
\lambda_{Zimm}=\frac{F[\eta]M_w\eta_s}{N_Ak_BT}
\label{equ:Zimm}
\end{equation}\
where $N_A$ is the Avogadro constant, $k_B$ is the Boltzmann constant (and the product $N_Ak_B$ being equal to the universal gas constant $R$), $T$ the absolute temperature and [$\eta$] is the intrinsic viscosity. Tirtaatmadja \al\cite{Tirt2006} have shown experimentally that this can be expressed as [$\eta$] = 0.072$M_w^{0.65}$ for the PEO solutions studied here (giving [$\eta$] in the usual units of ml$/$g).  The prefactor $F$ is given by Rodd \al \cite{Rodd2005} to be 0.463.  Rodd \al \cite{Rodd2007} measured the intrinsic viscosity for different polymer water/glycerol mixtures and have shown that it decreases due to a decrease in solvent quality. As a consequence the dependence of $\lambda$ on $\eta_s$ becomes less than linear. In Fig.\,\ref{fig:calibration}a) we show how these estimates of the Zimm relaxation time compare to those determined from the first normal-stress difference measured in the cone-and-plate rotational rheometer, illustrating that the agreement is very good.  Given the polydispersity inherent in such commercial polymers, and batch-to-batch variations which the data in Fig.\,\ref{fig:rheo_SB}b) highlights, the almost quantitative agreement between the two estimates of $\lambda$ is striking (especially given the various constants used in the determination of $\lambda_{Zimm}$ from equation \eqref{equ:Zimm}). Such agreement provides confidence in the robustness of our estimates of the relaxation time and hence in the calibration of the serpentine rheometer.

\section{Calibration of the serpentine rheometer}

To calibrate the serpentine rheometer the value of $C$ (eqn.\ \eqref{equ:geometric_scaling_visc}) has to be determined. To do so we compare results for $a=\lambda/C$ from the serpentine channel to relaxation times from classical rheometry. To obtain better accuracy, we not only compare these values for a single polymer solution, but also use solutions of PEO 2Mio for different solvent viscosities and different concentrations.

Firstly we determine the values of $a=\lambda/C$ from the slope obtained from the fits of the critical shear rate versus the radius of curvature (Fig.\,\ref{fig:fits_serpentine channel}) and the ratio of the polymer to the solvent viscosity for PEO 2Mio for two different concentrations 125\,ppm and 400\,ppm. These results are then compared to the results for $\lambda$ from the classical shear rheometer (see Fig.\,\ref{fig:rheo_SB}b)) for PEO 2Mio at a concentration of 400\,ppm. Note that it was not possible to obtain reliable measurements for the smaller concentration of 125\,ppm on the classical rheometer. Fig.\,\ref{fig:calibration}b) shows $\lambda$ together with $a$ as a function of the solvent viscosity $\eta_s$.

 A number of things should be remarked from Fig.\,\ref{fig:calibration}b). First, all three data sets show a comparable dependence of $\lambda$ on the solvent viscosity, that we will continue to describe as $\lambda \sim \eta_s^{0.9}$. Second, the results obtained from the serpentine rheometer for the two different concentrations are in good agreement. This shows that our correction for the solvent viscosity is adequate and is a first validation that the proposed rheometer works very well. We adjusted $a=\lambda/C=B\eta_s^{0.9}$ and obtained $B=4.99\pm0.23$\,ms/(mPa\,s)$^{0.9}$ as the best fit for both curves together. Finally by comparing $A=0.25\pm0.02$\,ms/(mPa\,s)$^{0.9}$ (see table \ref{tab:table}) from the classical rheometry to the value of $B$ from the serpentine rheometer we obtain $C=0.05$.

\section{Using the serpentine channel}

We now discuss possible applications of the serpentine rheometer.

\subsection{A quantitative rheometer}

\begin{figure*}[ht!]
	\centering
       \includegraphics[height=6.5cm]{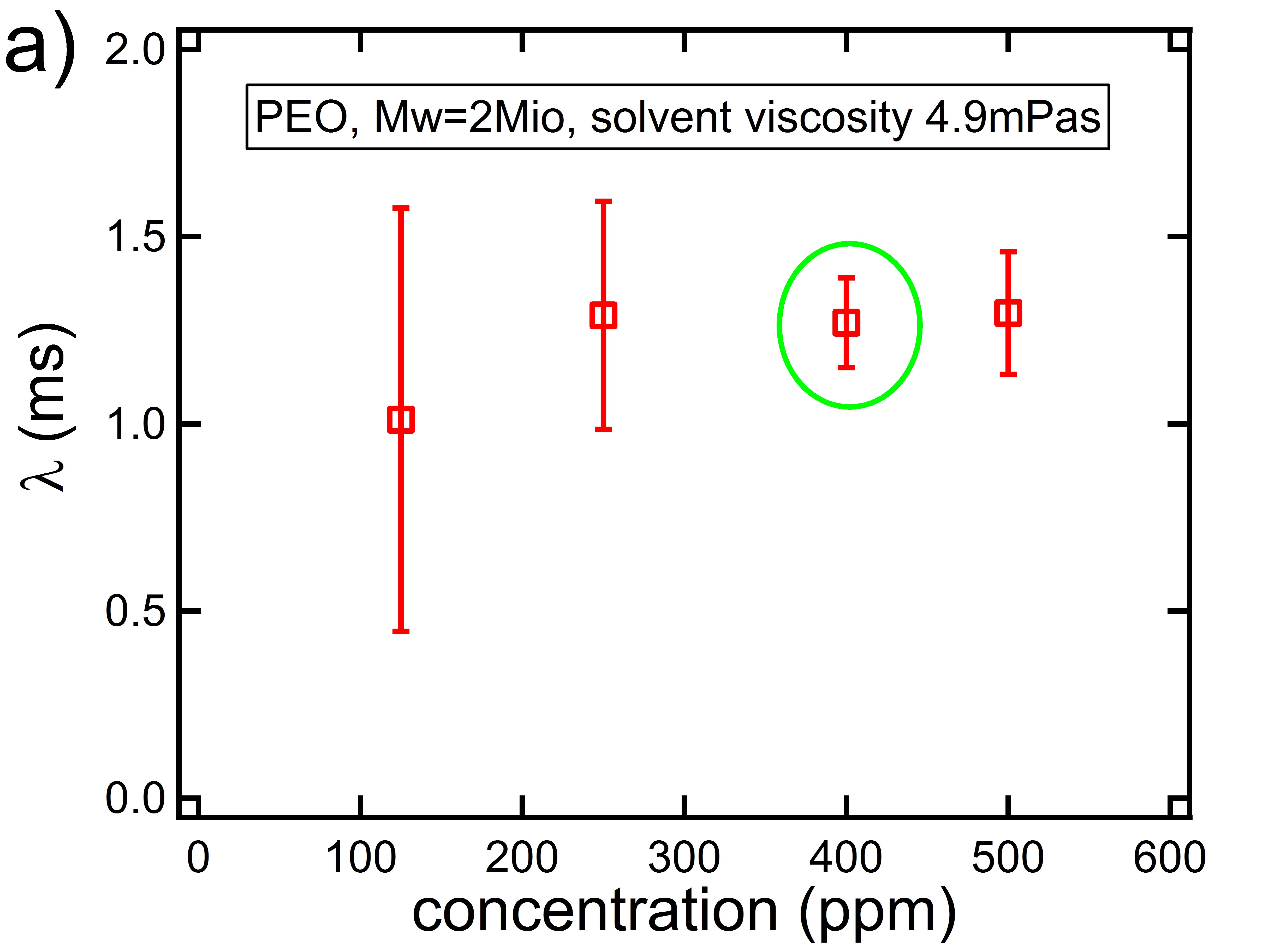}
       \includegraphics[height=6.5cm]{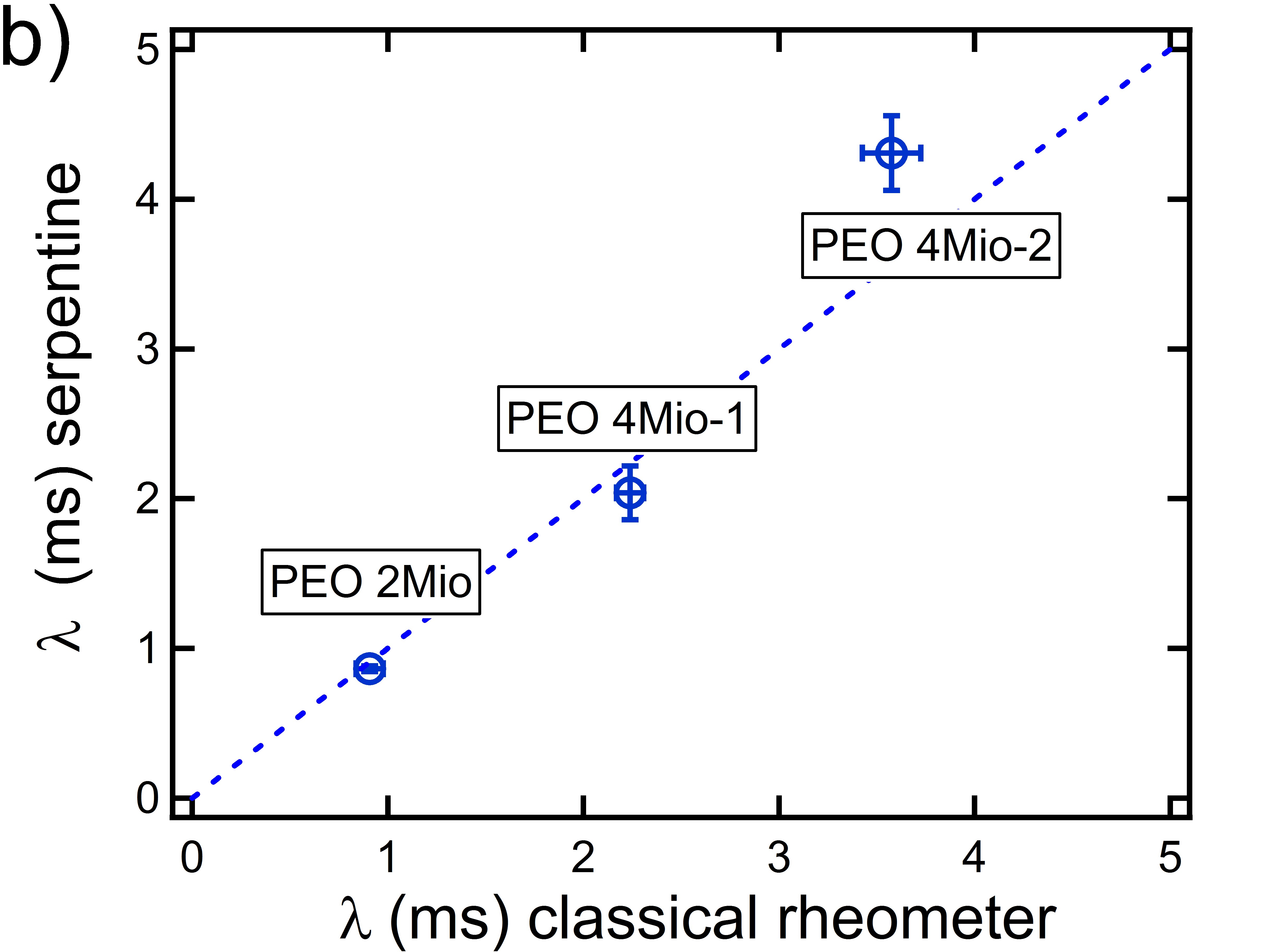}
	\caption{a) Relaxation time $\lambda$ obtained from the serpentine rheometer for PEO 2Mio in a solvent with viscosity $\eta_s=4.9$\,mPa\,s at varying concentration, ranging from 125\,ppm to 500\,ppm. The data points represent the average over three sets of experiments and the corresponding error bars. The relaxation time at 400\,ppm corresponds to the value from the calibration curve. b) Relaxation times from the serpentine rheometer versus relaxation times from classical shear rheometry. The results from the serpentine rheometer correspond to a concentration of 125\,ppm for the three different polymers and a solvent viscosity of $\eta_s=5.2$\,mPa\,s. The error bars are estimated from the error of the fit to $\s_c$. The results from the classical rheometer for 400\,ppm solutions are calculated from the fits of the relaxation time with the solvent viscosity, see Table \ref{tab:table}. In this case the error is estimated from the error of the fit on $A$.}
	\label{fig:serpentine_results}
\end{figure*}

\subsubsection{Extension to lower concentrations}

\ First we measure the relaxation times for different concentrations $c$ of a given polymer. As we are working in the dilute regime ($c<c^*$), for a given molecular weight we expect to obtain identical relaxation times. The results for the PEO 2Mio and concentrations $c=125$\,ppm, 250\,ppm and 500\,ppm for a solvent viscosity of $\eta_s=4.9$\,mPa\,s are presented in Fig.\,\ref{fig:serpentine_results}a). These results were obtained from an independent series of measurements and each experiment has been repeated three times using fresh polymer solutions. The results are compared to the value for the relaxation time at $c=400$\,ppm from the calibration curve (highlighted by a circle in Fig.\,\ref{fig:calibration}b). Note that we did not include the value for $c=125$\,ppm from the calibration curve as we did not perform experiments with the corresponding solvent viscosity for this concentration.

Not withstanding the rather large uncertainty for the smallest concentration, these results are very promising: a value of $\lambda\approx 1.2$\,ms is found independent of the polymer concentration. Note that it was not possible to measure the relaxation time of the low concentration $c=125$~ppm using the rotational rheometer, indicating the superior sensitivity of the serpentine channel, which is able to measure very small relaxation times down to very small concentrations. This is in agreement with the findings from the calibration curve that as long as the correction for the ratio between the solvent and the polymer viscosity is made correctly, identical results are obtained for different concentrations in the dilute regime. We remark that additional miniaturization of the serpentine channel enhances the elastic effects, thus increasing further the sensitivity of this rheometric device. A recent review \cite{GR2013} discusses the challenges of measuring viscoelastic properties of dilute polymer solutions, highlighting the relevance of using microfluidic devices, as in our work, to probe the elastic properties of such polymer solutions.

\subsubsection{Changing the molecular weight\;\;} Secondly we have measured the relaxation time for solutions of PEO of different molecular weight: 2Mio and two different batches of 4Mio, denoted 4Mio-1 and 4Mio-2. Fresh solutions for all three polymers were prepared at 125\,ppm and a solvent viscosity of $\eta_s=5.2$\,mPa\,s. The relaxation times obtained are compared to the relaxation times from classical shear rheometry measured at 400\,ppm. As we are in the dilute regime, no dependence of $\lambda$ with the polymer concentration is expected, and we have explicitly shown that this is true for the solution of PEO 2Mio in Fig.\,\ref{fig:serpentine_results}a). The values used from the classical rotational rheometry are calculated using the fit parameters from Table \ref{tab:table} to obtain the relaxation times at the solvent viscosity of $\eta_s=5.2$\,mPa\,s.

The results from the serpentine channel are plotted in Fig.\,\ref{fig:serpentine_results}b) against the results from the classical rheometry and satisfactory agreement between these two independent techniques is obtained. Note that the fact that the error bars are smaller than the differences between these two measurements in some cases is very likely due to the fact that independently prepared polymer solutions have been used in each measurement on the two different devices.

In addition, as was actually already apparent from the classical shear rheology data of Fig.\,\ref{fig:rheo_SB}b), we note that the data for the two batches of 4Mio PEO highlight large batch-to-batch variations that can occur between nominally-identical samples: there is a factor of two difference in relaxation time between both samples.  Such differences are likely a consequence of differing degrees of polydispersity in the two batches.

\subsection{Integration into a microfluidic lab-on-a-chip device}

As the foregoing makes clear, to determine a quantitative value of the relaxation time using the serpentine rheometer, an independent measurement of the polymer contribution to the total viscosity (and indeed a measurement of the solvent viscosity if it is unknown) is required such that the value of $\sqrt{\eta/\eta_p}$ can be determined to use in equation \eqref{equ:geometric_scaling_visc}.  In the current study these values were obtained from separate measurements using an Ubbelohde capillary viscometer.  Ideally one would like to be able to determine both the viscosity ratio and the critical shear rate from a single microfluidic lab-on-a-chip device.  To do so one could either use pressure drop measurements in a straight section upstream of the serpentine channel, as in Pipe \al \cite{Pipe2009} for example, or, alternatively, use the Y-junction approach of Guillot \al\cite{Guillot2008} or Nghe \al\cite{Nghe2010}.  As the easiest method to observe the purely-elastic instability in the serpentine channel is via an optical visualization technique, integration into a system based on the latter approach is probably to be preferred thereby avoiding the requirement of incorporating pressure sensors into the device.  To avoid issues of possible polymer degradation due to the instability, in the pressure-drop case the viscometer section of the device should be incorporated upstream of the serpentine channel and, in the Y-junction case where a reference fluid is required, in a separate parallel micro-channel on the same chip.

\subsection{Using the serpentine rheometer as a comparator or index device with an application to polymer degradation}

If one is not concerned with the absolute value of a fluid's relaxation time {\it per se} but rather with indexing different fluids according to their degree of elasticity, then the serpentine channel represents an extremely efficient device.  In this scenario it is simply sufficient to test the different fluids in a single channel of known curvature and determine the critical shear rate in each case.  Following equation \eqref{equ:geometric_scaling_visc} this directly leads to $\s_{c,1}/\s_{c,2}=\lambda_2/\lambda_1$.  Of course, strictly speaking, this equality only holds for fluids where the ratio $\eta/\eta_p$ remains constant.  In the absence of quantitative information regarding the contribution to polymer viscosity, a pragmatic approach, if the critical overlap concentration is known, is to use the scaling\cite{Larson1999} ${c^*[\eta]}$ ${\sim}$ 1 which gives  $\eta_p$ ${\sim}$ ${\eta_s}$ $c/{c^*}$, or $\eta/\eta_p=1/(1+c^*/c)$.

\begin{figure}[h]
	\centering
         \includegraphics[height=6.5cm]{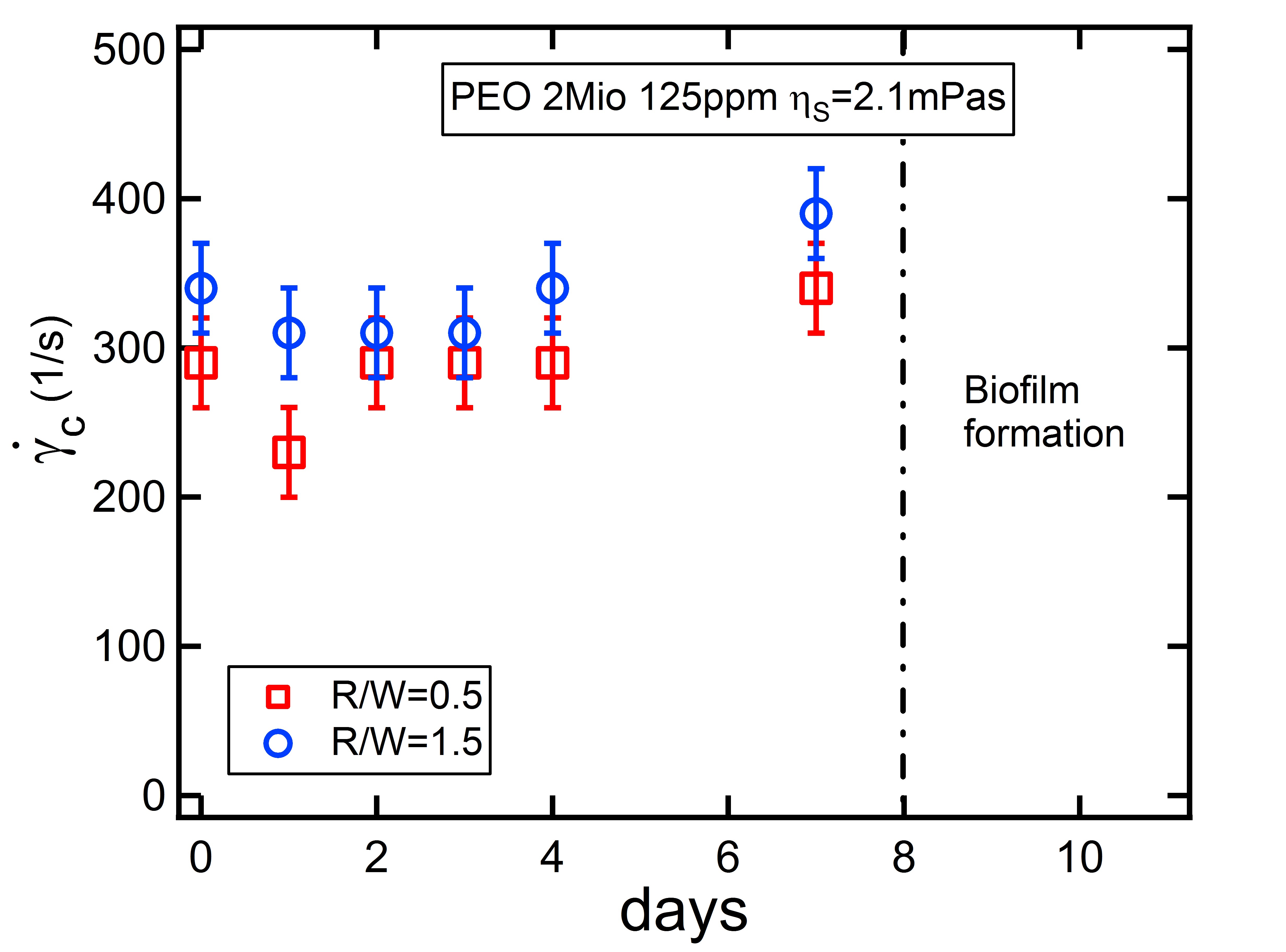}
	\caption{Critical shear rate measured over time using the serpentine rheometer for two different radii for a PEO sample exposed to sunlight. The error bars correspond to the uncertainty in measuring the critical flow rate.}
	\label{fig:degradation}
\end{figure}

 Alternatively the critical shear rate can be used for multiple repeat experiments of the same fluid to test for degradation.  Often one needs to check if polymer degradation has occurred due to either photo-induced, mechanical, thermal, chemical or biological causes\cite{Vana2005,Vana2006,Nghe2010}.  Simple shear viscosity measurements are often fairly insensitive to such effects as degraded polymers often still contribute to the overall viscosity of the solution which tends to be dominated by the solvent viscosity in any case for dilute polymer solutions (and is therefore a small effect). In contrast, the polymer relaxation time is a much more sensitive harbinger of degradation.  In Fig.\,\ref{fig:degradation} we show the effect of photo-induced degradation on a virgin polymer solution, i.e unsheared, stored at room temperature in a clear bottle exposed to sunlight over a period of several days.  Each day a new measurement was made and, after seven days, a precursive slight increase in the critical shear rate was observed which was followed by destruction of the sample via the formation of biofilms. Finally, our experience with the serpentine rheometer suggests its sensitivity makes it an ideal instrument for monitoring possible degradation effects regardless of the precise degrading mechanism.

\section{Conclusions}

By understanding the scaling behaviour of the onset of a purely-elastic flow instability in a microfluidic serpentine channel we have proposed a microfluidic rheometric device which is capable of measuring fluid relaxation times down to 1\,ms.  In contrast to most other rheometers, which aim to produce viscometric flows to enable the extraction of rheological properties, the device makes use of an inherent instability within the flow to estimate the fluid relaxation time.  Although using interfacial instabilities has previously been tentatively proposed \cite{Khom2000,Bon2011} as a means of estimating material properties as has the onset of viscoelastic vortices\cite{CD2011}, the current method, which only requires a single fluid, is the first to show that relaxation times can be successfully measured using such an approach.  Also, very recently, Koser \al\cite{koser2013} have proposed using creep-recovery tests in a microfluidic device to estimate polymer relaxation times.  However, their device requires the use of a high-speed camera and is restricted to relaxation times at least an order of magnitude greater than those measured here. The serpentine rheometer can either; (a) be used as a comparator or indexing device in which case fluids can be ranked according to their elasticity or changes monitored, such as those due to degradation or (b) be used as a true rheometer once calibration tests, using classical cone-and-plate rheometry for example, have been performed. In this latter case, the microfluidic serpentine device can access lower molecular weight materials, solvent viscosities or concentrations than is possible using state-of-the-art commercial rheometers.

\section*{Acknowledgements}
Some of this work was undertaken whilst RJP was a visiting "Chaire Michelinâ" at the ESPCI Paris Tech in June 2013 and this support is hereby gratefully acknowledged.  MAA acknowledges funding from the European Research Council (ERC), under the European Commission "Ideas" Specific Programme of the Seventh Framework Programme (Grant Agreement no 307499).

\footnotesize{
\providecommand*{\mcitethebibliography}{\thebibliography}
\csname @ifundefined\endcsname{endmcitethebibliography}
{\let\endmcitethebibliography\endthebibliography}{}

\bibliographystyle{rsc} 

\begin{mcitethebibliography}{38}
\providecommand*{\natexlab}[1]{#1}
\providecommand*{\mciteSetBstSublistMode}[1]{}
\providecommand*{\mciteSetBstMaxWidthForm}[2]{}
\providecommand*{\mciteBstWouldAddEndPuncttrue}
  {\def\EndOfBibitem{\unskip.}}
\providecommand*{\mciteBstWouldAddEndPunctfalse}
  {\let\EndOfBibitem\relax}
\providecommand*{\mciteSetBstMidEndSepPunct}[3]{}
\providecommand*{\mciteSetBstSublistLabelBeginEnd}[3]{}
\providecommand*{\EndOfBibitem}{}
\mciteSetBstSublistMode{f}
\mciteSetBstMaxWidthForm{subitem}
{(\emph{\alph{mcitesubitemcount}})}
\mciteSetBstSublistLabelBeginEnd{\mcitemaxwidthsubitemform\space}
{\relax}{\relax}

\bibitem[Bird \emph{et~al.}(1987)Bird, Armstrong, and Hassager]{Bird1987}
R.~Bird, R.~Armstrong and O.~Hassager, \emph{{Dynamics of polymeric liquids}},
  1987\relax
\mciteBstWouldAddEndPuncttrue
\mciteSetBstMidEndSepPunct{\mcitedefaultmidpunct}
{\mcitedefaultendpunct}{\mcitedefaultseppunct}\relax
\EndOfBibitem
\bibitem[Virk(1975)]{Virk1975}
P.~S. Virk, \emph{AIChE Journal}, 1975, \textbf{21}, 625--656\relax
\mciteBstWouldAddEndPuncttrue
\mciteSetBstMidEndSepPunct{\mcitedefaultmidpunct}
{\mcitedefaultendpunct}{\mcitedefaultseppunct}\relax
\EndOfBibitem
\bibitem[Jones and Walters(1989)]{Jones1989}
D.~Jones and K.~Walters, \emph{Rheologica Acta}, 1989, \textbf{28},
  482--498\relax
\mciteBstWouldAddEndPuncttrue
\mciteSetBstMidEndSepPunct{\mcitedefaultmidpunct}
{\mcitedefaultendpunct}{\mcitedefaultseppunct}\relax
\EndOfBibitem
\bibitem[Lindner \emph{et~al.}(2003)Lindner, Vermant, and Bonn]{Lindner2003}
A.~Lindner, J.~Vermant and D.~Bonn, \emph{Physica A: Statistical Mechanics and
  its Applications}, 2003, \textbf{319}, 125--133\relax
\mciteBstWouldAddEndPuncttrue
\mciteSetBstMidEndSepPunct{\mcitedefaultmidpunct}
{\mcitedefaultendpunct}{\mcitedefaultseppunct}\relax
\EndOfBibitem
\bibitem[Guillot \emph{et~al.}({2006})Guillot, Panizza, Salmon, Joanicot,
  Colin, Bruneau, and Colin]{Guillot2006}
P.~Guillot, P.~Panizza, J.-B. Salmon, M.~Joanicot, A.~Colin, C.-H. Bruneau and
  T.~Colin, \emph{{Langmuir}}, {2006}, \textbf{{22}}, {6438--6445}\relax
\mciteBstWouldAddEndPuncttrue
\mciteSetBstMidEndSepPunct{\mcitedefaultmidpunct}
{\mcitedefaultendpunct}{\mcitedefaultseppunct}\relax
\EndOfBibitem
\bibitem[Guillot \emph{et~al.}(2008)Guillot, Moulin, K\"{o}titz, Guirardel,
  Dodge, Joanicot, Colin, Bruneau, and Colin]{Guillot2008}
P.~Guillot, T.~Moulin, R.~K\"{o}titz, M.~Guirardel, A.~Dodge, M.~Joanicot,
  A.~Colin, C.-H. Bruneau and T.~Colin, \emph{Microfluidics and Nanofluidics},
  2008, \textbf{5}, 619--630\relax
\mciteBstWouldAddEndPuncttrue
\mciteSetBstMidEndSepPunct{\mcitedefaultmidpunct}
{\mcitedefaultendpunct}{\mcitedefaultseppunct}\relax
\EndOfBibitem
\bibitem[Pipe \emph{et~al.}({2008})Pipe, Majmudar, and McKinley]{Pipe2008}
C.~J. Pipe, T.~S. Majmudar and G.~H. McKinley, \emph{{Rheologica Acta}},
  {2008}, \textbf{{47}}, {621--642}\relax
\mciteBstWouldAddEndPuncttrue
\mciteSetBstMidEndSepPunct{\mcitedefaultmidpunct}
{\mcitedefaultendpunct}{\mcitedefaultseppunct}\relax
\EndOfBibitem
\bibitem[Chevalier and Ayela({2008})]{Chev2008}
J.~Chevalier and F.~Ayela, \emph{{Review of Scientific Instruments}}, {2008},
  {076102}\relax
\mciteBstWouldAddEndPuncttrue
\mciteSetBstMidEndSepPunct{\mcitedefaultmidpunct}
{\mcitedefaultendpunct}{\mcitedefaultseppunct}\relax
\EndOfBibitem
\bibitem[Nghe \emph{et~al.}({2010})Nghe, Tabeling, and Ajdari]{Nghe2010}
P.~Nghe, P.~Tabeling and A.~Ajdari, \emph{{Journal of Non-Newtonian Fluid
  Mechanics}}, {2010}, \textbf{{165}}, {313--322}\relax
\mciteBstWouldAddEndPuncttrue
\mciteSetBstMidEndSepPunct{\mcitedefaultmidpunct}
{\mcitedefaultendpunct}{\mcitedefaultseppunct}\relax
\EndOfBibitem
\bibitem[Livak-Dahl \emph{et~al.}({2013})Livak-Dahl, Lee, and Burns]{Livak2013}
E.~Livak-Dahl, J.~Lee and M.~A. Burns, \emph{{Lab on a Chip}}, {2013},
  \textbf{{13}}, {297--301}\relax
\mciteBstWouldAddEndPuncttrue
\mciteSetBstMidEndSepPunct{\mcitedefaultmidpunct}
{\mcitedefaultendpunct}{\mcitedefaultseppunct}\relax
\EndOfBibitem
\bibitem[Berthet \emph{et~al.}(2011)Berthet, Jundt, Durivault, Mercier, and
  Angelescu]{Berthet2011}
H.~Berthet, J.~Jundt, J.~Durivault, B.~Mercier and D.~Angelescu, \emph{Lab on a
  chip}, 2011, \textbf{11}, 215--23\relax
\mciteBstWouldAddEndPuncttrue
\mciteSetBstMidEndSepPunct{\mcitedefaultmidpunct}
{\mcitedefaultendpunct}{\mcitedefaultseppunct}\relax
\EndOfBibitem
\bibitem[Pipe and McKinley({2009})]{Pipe2009}
C.~J. Pipe and G.~H. McKinley, \emph{{Mechanics Research Communications}},
  {2009}, \textbf{{36}}, {110--120}\relax
\mciteBstWouldAddEndPuncttrue
\mciteSetBstMidEndSepPunct{\mcitedefaultmidpunct}
{\mcitedefaultendpunct}{\mcitedefaultseppunct}\relax
\EndOfBibitem
\bibitem[Oliveira \emph{et~al.}(2008)Oliveira, Rodd, McKinley, and
  Alves]{Oliveira2008}
M.~S.~N. Oliveira, L.~E. Rodd, G.~H. McKinley and M.~A. Alves, \emph{Microfluid
  Nanofluid}, 2008, \textbf{5}, 809--826\relax
\mciteBstWouldAddEndPuncttrue
\mciteSetBstMidEndSepPunct{\mcitedefaultmidpunct}
{\mcitedefaultendpunct}{\mcitedefaultseppunct}\relax
\EndOfBibitem
\bibitem[Nelson \emph{et~al.}({2011})Nelson, Kavehpour, and Kim]{Nelson2011}
W.~C. Nelson, H.~P. Kavehpour and C.-J.~C. Kim, \emph{{Lab on a Chip}}, {2011},
  \textbf{{11}}, {2424--2431}\relax
\mciteBstWouldAddEndPuncttrue
\mciteSetBstMidEndSepPunct{\mcitedefaultmidpunct}
{\mcitedefaultendpunct}{\mcitedefaultseppunct}\relax
\EndOfBibitem
\bibitem[Haward \emph{et~al.}(2012)Haward, Oliveira, Alves, and
  McKinley]{Haward2012}
S.~J. Haward, M.~S.~N. Oliveira, M.~A. Alves and G.~H. McKinley, \emph{Physical
  Review Letters}, 2012, \textbf{109}, 128301\relax
\mciteBstWouldAddEndPuncttrue
\mciteSetBstMidEndSepPunct{\mcitedefaultmidpunct}
{\mcitedefaultendpunct}{\mcitedefaultseppunct}\relax
\EndOfBibitem
\bibitem[Galindo-Rosales \emph{et~al.}({2013})Galindo-Rosales, Alves, and
  Oliveira]{GR2013}
F.~J. Galindo-Rosales, M.~A. Alves and M.~S.~N. Oliveira, \emph{{Microfluidics
  and Nanofluidics}}, {2013}, \textbf{{14}}, {1--19}\relax
\mciteBstWouldAddEndPuncttrue
\mciteSetBstMidEndSepPunct{\mcitedefaultmidpunct}
{\mcitedefaultendpunct}{\mcitedefaultseppunct}\relax
\EndOfBibitem
\bibitem[Christopher \emph{et~al.}({2010})Christopher, Yoo, Dagalakis, Hudson,
  and Migler]{Chris2010}
G.~F. Christopher, J.~M. Yoo, N.~Dagalakis, S.~D. Hudson and K.~B. Migler,
  \emph{{Lab on a Chip}}, {2010}, \textbf{{10}}, {2749--2757}\relax
\mciteBstWouldAddEndPuncttrue
\mciteSetBstMidEndSepPunct{\mcitedefaultmidpunct}
{\mcitedefaultendpunct}{\mcitedefaultseppunct}\relax
\EndOfBibitem
\bibitem[Larson \emph{et~al.}(1990)Larson, Shaqfeh, and Muller]{Larson1990}
R.~G. Larson, E.~S.~G. Shaqfeh and S.~J. Muller, \emph{Journal of Fluid
  Mechanics}, 1990, \textbf{218}, 573\relax
\mciteBstWouldAddEndPuncttrue
\mciteSetBstMidEndSepPunct{\mcitedefaultmidpunct}
{\mcitedefaultendpunct}{\mcitedefaultseppunct}\relax
\EndOfBibitem
\bibitem[Groisman and Steinberg(2004)]{Groisman2004a}
A.~Groisman and V.~Steinberg, \emph{New Journal of Physics}, 2004, \textbf{6},
  29--29\relax
\mciteBstWouldAddEndPuncttrue
\mciteSetBstMidEndSepPunct{\mcitedefaultmidpunct}
{\mcitedefaultendpunct}{\mcitedefaultseppunct}\relax
\EndOfBibitem
\bibitem[Shaqfeh(1996)]{Shaqfeh1996}
E.~S.~G. Shaqfeh, \emph{Annual Review of Fluid Mechanics}, 1996, \textbf{28},
  129--85\relax
\mciteBstWouldAddEndPuncttrue
\mciteSetBstMidEndSepPunct{\mcitedefaultmidpunct}
{\mcitedefaultendpunct}{\mcitedefaultseppunct}\relax
\EndOfBibitem
\bibitem[McKinley \emph{et~al.}(1996)McKinley, Pakdel, and
  Oztekin]{McKinley1996}
G.~McKinley, P.~Pakdel and A.~Oztekin, \emph{Journal of Non-Newtonian Fluid
  Mechanics}, 1996, \textbf{67}, 19--47\relax
\mciteBstWouldAddEndPuncttrue
\mciteSetBstMidEndSepPunct{\mcitedefaultmidpunct}
{\mcitedefaultendpunct}{\mcitedefaultseppunct}\relax
\EndOfBibitem
\bibitem[Khomami and Su({2000})]{Khom2000}
B.~Khomami and K.~Su, \emph{{Journal of Non-Newtonian Fluid Mechanics}},
  {2000}, \textbf{{91}}, {59--84}\relax
\mciteBstWouldAddEndPuncttrue
\mciteSetBstMidEndSepPunct{\mcitedefaultmidpunct}
{\mcitedefaultendpunct}{\mcitedefaultseppunct}\relax
\EndOfBibitem
\bibitem[Groisman and Quake({2004})]{Grois2004}
A.~Groisman and S.~Quake, \emph{{Physical Review Letters}}, {2004},
  \textbf{{92}}, {094501}\relax
\mciteBstWouldAddEndPuncttrue
\mciteSetBstMidEndSepPunct{\mcitedefaultmidpunct}
{\mcitedefaultendpunct}{\mcitedefaultseppunct}\relax
\EndOfBibitem
\bibitem[Poole \emph{et~al.}({2007})Poole, Alves, and Oliveira]{Poole2007}
R.~J. Poole, M.~A. Alves and P.~J. Oliveira, \emph{{Physical Review Letters}},
  {2007}, \textbf{{99}}, {164503}\relax
\mciteBstWouldAddEndPuncttrue
\mciteSetBstMidEndSepPunct{\mcitedefaultmidpunct}
{\mcitedefaultendpunct}{\mcitedefaultseppunct}\relax
\EndOfBibitem
\bibitem[Morozov and van Saarloos({2007})]{Morozov2007}
A.~N. Morozov and W.~van Saarloos, \emph{{Physics Reports - Review Section of
  Physics Letters}}, {2007}, \textbf{{447}}, {112--143}\relax
\mciteBstWouldAddEndPuncttrue
\mciteSetBstMidEndSepPunct{\mcitedefaultmidpunct}
{\mcitedefaultendpunct}{\mcitedefaultseppunct}\relax
\EndOfBibitem
\bibitem[Bonhomme \emph{et~al.}({2011})Bonhomme, Morozov, Leng, and
  Colin]{Bon2011}
O.~Bonhomme, A.~Morozov, J.~Leng and A.~Colin, \emph{{Physical Review E}},
  {2011}, \textbf{{83}}, {065301}\relax
\mciteBstWouldAddEndPuncttrue
\mciteSetBstMidEndSepPunct{\mcitedefaultmidpunct}
{\mcitedefaultendpunct}{\mcitedefaultseppunct}\relax
\EndOfBibitem
\bibitem[Pakdel and McKinley(1996)]{Pakdel1996}
P.~Pakdel and G.~McKinley, \emph{Physical Review Letters}, 1996, \textbf{77},
  2459--2462\relax
\mciteBstWouldAddEndPuncttrue
\mciteSetBstMidEndSepPunct{\mcitedefaultmidpunct}
{\mcitedefaultendpunct}{\mcitedefaultseppunct}\relax
\EndOfBibitem
\bibitem[Zilz \emph{et~al.}(2012)Zilz, Poole, Alves, Bartolo, Levach\'{e}, and
  Lindner]{Zilz2012}
J.~Zilz, R.~J. Poole, M.~A. Alves, D.~Bartolo, B.~Levach\'{e} and A.~Lindner,
  \emph{Journal of Fluid Mechanics}, 2012, \textbf{712}, 203--218\relax
\mciteBstWouldAddEndPuncttrue
\mciteSetBstMidEndSepPunct{\mcitedefaultmidpunct}
{\mcitedefaultendpunct}{\mcitedefaultseppunct}\relax
\EndOfBibitem
\bibitem[Poole \emph{et~al.}(2013)Poole, Lindner, and Alves]{Poole2013}
R.~J. Poole, A.~Lindner and M.~A. Alves, \emph{Journal of Non-Newtonian Fluid
  Mechanics}, 2013, \textbf{201}, 10--16\relax
\mciteBstWouldAddEndPuncttrue
\mciteSetBstMidEndSepPunct{\mcitedefaultmidpunct}
{\mcitedefaultendpunct}{\mcitedefaultseppunct}\relax
\EndOfBibitem
\bibitem[Rodd \emph{et~al.}(2007)Rodd, Cooper-White, Boger, and
  McKinley]{Rodd2007}
L.~Rodd, J.~Cooper-White, D.~Boger and G.~McKinley, \emph{Journal of
  Non-Newtonian Fluid Mechanics}, 2007, \textbf{143}, 170--191\relax
\mciteBstWouldAddEndPuncttrue
\mciteSetBstMidEndSepPunct{\mcitedefaultmidpunct}
{\mcitedefaultendpunct}{\mcitedefaultseppunct}\relax
\EndOfBibitem
\bibitem[Zell \emph{et~al.}(2010)Zell, Gier, Rafa, and Wagner]{Zell2010}
A.~Zell, S.~Gier, S.~Rafa and C.~Wagner, \emph{Journal of Non-Newtonian Fluid
  Mechanics}, 2010, \textbf{165(19)}, 1--24\relax
\mciteBstWouldAddEndPuncttrue
\mciteSetBstMidEndSepPunct{\mcitedefaultmidpunct}
{\mcitedefaultendpunct}{\mcitedefaultseppunct}\relax
\EndOfBibitem
\bibitem[Larson(1999)]{Larson1999}
R.~G. Larson, \emph{{The Structure and Rheology of Complex Fluids}}, 1999\relax
\mciteBstWouldAddEndPuncttrue
\mciteSetBstMidEndSepPunct{\mcitedefaultmidpunct}
{\mcitedefaultendpunct}{\mcitedefaultseppunct}\relax
\EndOfBibitem
\bibitem[Tirtaatmadja \emph{et~al.}({2006})Tirtaatmadja, McKinley, and
  Cooper-White]{Tirt2006}
V.~Tirtaatmadja, G.~McKinley and J.~Cooper-White, \emph{{Physics of Fluids}},
  {2006}, \textbf{{18}}, {043101}\relax
\mciteBstWouldAddEndPuncttrue
\mciteSetBstMidEndSepPunct{\mcitedefaultmidpunct}
{\mcitedefaultendpunct}{\mcitedefaultseppunct}\relax
\EndOfBibitem
\bibitem[Rodd \emph{et~al.}(2005)Rodd, Scott, Boger, Cooper-White, and
  McKinley]{Rodd2005}
L.~Rodd, T.~Scott, D.~Boger, J.~Cooper-White and G.~McKinley, \emph{Journal of
  Non-Newtonian Fluid Mechanics}, 2005, \textbf{129}, 1--22\relax
\mciteBstWouldAddEndPuncttrue
\mciteSetBstMidEndSepPunct{\mcitedefaultmidpunct}
{\mcitedefaultendpunct}{\mcitedefaultseppunct}\relax
\EndOfBibitem
\bibitem[Vanapalli \emph{et~al.}({2005})Vanapalli, Islam, and
  Solomon]{Vana2005}
S.~Vanapalli, M.~Islam and M.~Solomon, \emph{{Physics of Fluids}}, {2005},
  \textbf{{17}}, {095108}\relax
\mciteBstWouldAddEndPuncttrue
\mciteSetBstMidEndSepPunct{\mcitedefaultmidpunct}
{\mcitedefaultendpunct}{\mcitedefaultseppunct}\relax
\EndOfBibitem
\bibitem[Vanapalli \emph{et~al.}({2006})Vanapalli, Ceccio, and
  Solomon]{Vana2006}
S.~A. Vanapalli, S.~L. Ceccio and M.~J. Solomon, \emph{{Proceedings of the
  National Academy of Sciences of the United States of America}}, {2006},
  \textbf{{103}}, {16660--16665}\relax
\mciteBstWouldAddEndPuncttrue
\mciteSetBstMidEndSepPunct{\mcitedefaultmidpunct}
{\mcitedefaultendpunct}{\mcitedefaultseppunct}\relax
\EndOfBibitem
\bibitem[Campo-Deano \emph{et~al.}({2011})Campo-Deano, Galindo-Rosales, Pinho,
  Alves, and Oliveira]{CD2011}
L.~Campo-Deano, F.~J. Galindo-Rosales, F.~T. Pinho, M.~A. Alves and M.~S.~N.
  Oliveira, \emph{{Journal of Non-Newtonian Fluid Mechanics}}, {2011},
  \textbf{{166}}, {1286--1296}\relax
\mciteBstWouldAddEndPuncttrue
\mciteSetBstMidEndSepPunct{\mcitedefaultmidpunct}
{\mcitedefaultendpunct}{\mcitedefaultseppunct}\relax
\EndOfBibitem
\bibitem[Koser \emph{et~al.}({2013})Koser, Pan, Keim, and Arratia]{koser2013}
A.~E. Koser, L.~Pan, N.~C. Keim and P.~E. Arratia, \emph{{Lab on a Chip}},
  {2013}, \textbf{{13}}, {1850--1853}\relax
\mciteBstWouldAddEndPuncttrue
\mciteSetBstMidEndSepPunct{\mcitedefaultmidpunct}
{\mcitedefaultendpunct}{\mcitedefaultseppunct}\relax
\EndOfBibitem
\end{mcitethebibliography}
}

\end{document}